\documentclass[a4paper,12pt]{article}
\usepackage[left=2cm,right=2cm,top=2cm,bottom=2cm]{geometry}
\usepackage[hang,bf,footnotesize]{caption}
\usepackage{graphicx}
\usepackage{subfigure}
\usepackage{xcolor}
\usepackage{amsmath}
\usepackage{mathptmx}
\usepackage{slashed}
\usepackage{cite}
\usepackage{bm}
\usepackage[utf8]{inputenc}

\allowdisplaybreaks[1]
\newcommand{\GeV}{\,\mathrm{GeV}}
\newcommand{\MeV}{\,\mathrm{MeV}}

\newcommand{\Lag}{\mathcal{L}}
\newcommand{\LSM}{\mathrm{L}\sigma\mathrm{M}}
\newcommand{\LLSM}{\mathcal{L}_{\mathrm{L}\sigma\mathrm{M}}}

\newcommand{\ie}{\textit{i.e.\,}}
\newcommand{\eg}{\textit{e.g.\,}}
\newcommand{\cf}{\textit{cf.\,}}

\newcommand{\Ord}[1]{\mathcal{O}(#1)}

\newcommand{\vpi}{\boldsymbol{\pi}}

\begin{document}
\begin{center}
   \textbf{\Large Imprints of a critical point on photon emission}\\[8mm]
   \begin{minipage}{0.9\textwidth}
      \begin{flushleft}
         F. Wunderlich$^{1,2}$ and B. K\"ampfer$^{1,2}$\\
         \textit{$^1$ \small Helmholtz-Zentrum Dresden-Rossendorf, D-01328 Dresden, Germany\\
                 $^2$ Technische Universit\"at Dresden, D-01062 Dresden, Germany}
      \end{flushleft}
   \end{minipage}
\end{center}

\abstract{The linear sigma model with linearized fluctuations of all involved fields facilitates the onset of a 
          sequence of first-order phase transitions at a critical point. This phase structure has distinctive imprints
          on the photon emission rates. We argue that analogously a critical point in the QCD phase diagram manifests
          itself by peculiarities of the photon spectra, in particular  when the dynamical expansion path of matter 
          crosses the phase transition curve in the vicinity of the critical point.\\[5mm]
} 
\section{Introduction}
The conjecture that QCD displays a non-trivial phase diagram has triggered a series of dedicated investigations.
On the experimental side, the beam energy scan at RHIC has brought up some remarkable results 
\cite{McDonald:2015tza} which are
under discussion w.r.t. an interpretation in terms of the onset of deconfinement at lower beam energies.
It is supported by earlier observations of particular features seen in the beam energy dependence of selected observables,
most notably the "horn" in the $K^+/\pi^+$ ratio \cite{Afanasiev:2002mx}. On the theoretical side, lattice QCD thermodynamic quantities 
have proven that the transition to deconfined matter at zero chemical potential occurs in a (rapid) crossover.
What remains unsettled is the existence of a first-order phase transition curve $T_c(\mu)$ in the temperature ($T$) vs.
chemical potential ($\mu$) plane, terminating in a (critical) end point (CEP) at $\mu_c>0$. The existence and the location
of the CEP, quantified by $T_c(\mu_c)$ and $\mu_c$, is a challenging task. 
With present techniques, ab initio QCD approaches can hardly fix the values of $T_c(\mu_c)$ and $\mu_c$ 
as well as the confinement-deconfinement
delineation curve $T_c(\mu)$ for $\mu>\mu_c$. Various models give widespread results \cite{Stephanov:2004wx,Friman:2011zz}.

The planned facilities NICA and FAIR will operate in a range of beam energies which is considered promising for 
the search of a QCD CEP. For example, the energy range, where the maximum freeze-out densities 
are achieved \cite{Randrup:2006nr,Randrup:2009ch}, will be covered by NICA;
it is also just the energy $\sqrt{s_{NN}}=\Ord{8\GeV}$ where the above mentioned "horn" is seen. Since NICA will
be equipped with detector installations which enable the measurement of photons, one may ask whether the direct photon
spectra have distinctive imprints of a CEP or the emerging first-order phase transition sequence.

To attempt an answer to this question, we are going to employ a particular
model with a CEP and calculate, within a kinetic theory approach, a few relevant channels for photon emission of 
strongly interacting matter in the vicinity of the CEP.

Owing to their penetrating nature, electromagnetic probes (dileptons as well as real photons) are promising tools 
for investigating the whole evolution of matter in the course of heavy-ion collisions,
although the convolution of signals from all stages of the evolution requires some effort in disentangling
the respective sources the photons. Especially interesting are the thermal photons 
from the hydrodynamic expansion stage if the system 
crosses the phase border line \cite{Bratkovskaya:2004kv,Ivanov:2005yw}. Thus,
understanding the experimental data (see \eg \cite{Tserruya:2009zt} for a review) requires precise theoretical
insight into the production processes (see \cite{Bauchle:2010ym,Linnyk:2013wma} for transport, 
\cite{Liu:2007zzw,vanHees:2011vb} 
for kinetic theory approaches to the thermal photon rates or \cite{Gale:2009gc,Rapp:2009yu} for recent reviews).

\section{A model with CEP: The linear sigma model}
The linear sigma model ($\LSM$) \cite{Scavenius:2000qd,Schaefer:2004en,Schaefer:2007pw,Herbst:2010rf} 
is known to display a CEP as endpoint of a first-order phase transition curve $T_c(\mu)$.
The degrees of freedom are quarks ($q=(u,d)$) and mesons ($\sigma, \pi^{0,\pm}$).
The Lagrangian reads 
\begin{eqnarray}
   \LLSM              &=& \bar q (i\slashed \partial  - g (\sigma + i\gamma_5\bm \tau \bm \pi))q - \Lag_{km}- U(\sigma, \bm \pi)\label{LSM_01},\\
   U(\sigma, \bm\pi) &=& \frac{\lambda}{4}(\sigma^2 + \bm \pi^2-\zeta)^2 - H\sigma\label{LSM_02},\\
   \Lag_{km}          &=& \frac{1}{2}(\partial_\mu\sigma\partial^\mu\sigma + \partial_\mu \bm\pi \partial^\mu \bm\pi)\label{LSM_03},
\end{eqnarray}
where $g$, $\zeta$, $\lambda$ and $H$ are parameters which can be  fixed, \eg by adjustments
to nucleon, pion and sigma masses as well as the pion decay constant. For convenience we choose the same parameter fixing 
as \cite{Mocsy:2004ab,Bowman:2008kc}, \ie  
$g=3.387$, $\zeta=7874\MeV^2$, $\lambda=27.58$ and $H=1.760\times10^6\MeV^3$.

The grand potential is given by a path integral of the exponential of the action constructed from $\LLSM$.
In mean field approximation (MFA) for the meson fields, one replaces the meson fields by their thermal averages, 
which results in the thermodynamic potential
\begin{eqnarray}
   \Omega_\text{MFA} &=& \Omega_{qq}\Big|_{m_q=gv} + U(v,\bm 0),
\end{eqnarray}
with $\Omega_{qq}$ being the potential constructed from the fermionic part of the Lagrangian and $v$ being the 
thermal expectation value of the $\sigma$ field. 
The resulting phase structure based on $\Omega_\text{MFA}$ is discussed in \cite{Scavenius:2000qd}.
Most important for our purpose is the variation of the effective masses over the phase diagram which is 
not restricted to the $\LSM$ but can also be found for other chiral models such as the NJL model 
\cite{Blaschke:2013zaa, Scavenius:2000qd}.

We include here, similar to \cite{Mocsy:2004ab,Bowman:2008kc,Ferroni:2010ct}, linearized fluctuations to account for quanta
of the meson fields which participate in reactions with photons in the exit channel but exclude vacuum terms for all the 
fields. 
In the linearized fluctuation approach (LFA) the fermionic part $\Omega_{qq}$ is calculated first and contributes, via the 
quark mass $m_q=g\sqrt{\sigma^2+\pi^2}$,
to the effective meson potential. This effective potential is afterwards approximated
by a parabola leading eventually to
\begin{equation}
   \begin{aligned}
      \Omega_\text{LFA} =&   \langle U(v+\Delta, \bm\pi)\rangle + \langle \Omega_{qq}(m_q) \rangle \\
                         &\quad - \frac12 m_\sigma^2\langle\Delta^2\rangle 
                                - \frac12 m_\pi^2\langle\bm\pi^2\rangle
                                + \Omega_\pi + \Omega_\sigma,
   \end{aligned}
\end{equation}
where $\sigma = v + \Delta$ and the brackets $\langle f(\sigma,\bm\pi)\rangle$ denoting the 
ensemble average of a function $f$ over $\sigma$ and $\pi$ fields and $\Omega_{\sigma,\pi}$ denoting the potential
for noninteracting bosons with effective masses $m_{\sigma,\pi}$ and multiplicities 1 and 3, respectively.
The emerging phase structure based on $\Omega_\text{LFA}$ is discussed in \cite{Bowman:2008kc,Ferroni:2010ct,Wunderlich:2015rwa}
(for full account of fluctuations, \cf \cite{Tripolt:2013jra}), where also the relations to symmetries of two-flavor QCD
are recollected.

\begin{figure}
   \centering
   \includegraphics[width=0.48\textwidth]{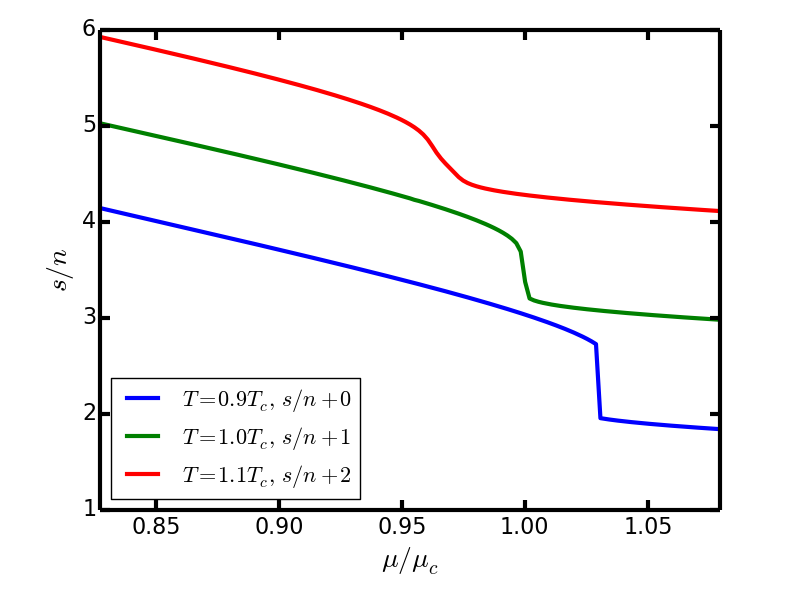}
   \caption{Entropy per quark $s/n$ as a function of scaled chemical potential for the $\LSM$ in LFA. 
            The ratio is shown for $T>T_c$, $T\approx T_c$ and $T<T_c$ (from top to bottom).
            For better visibility the upper curves have offsets of 1 and 2. 
            Parameters as in \cite{Wunderlich:2014cia,Wunderlich:2015rwa} yielding $\mu_c= 278\MeV$ and 
            $T_c\equiv T_c(\mu_c) = 74\MeV$, quite different from values advocated recently in \cite{Lacey:2014wqa} or 
            \cite{Tripolt:2013jra} which are above or below that temperature value.
            }
   \label{fig_entrop_per_baryon}
\end{figure}

The chiral transition of the $\LSM$ belongs to the class of liquid-gas type 
phase transitions \cite{Steinheimer:2013xxa}, where the entropy
per quark $s/n$ displays at the phase boundary a drop both with increasing temperature at constant 
chemical potential $\mu>\mu_c$ or with increasing chemical potential at constant temperature, see 
Fig.~\ref{fig_entrop_per_baryon}.

\begin{figure}
   \centering
   \includegraphics[width=0.48\textwidth]{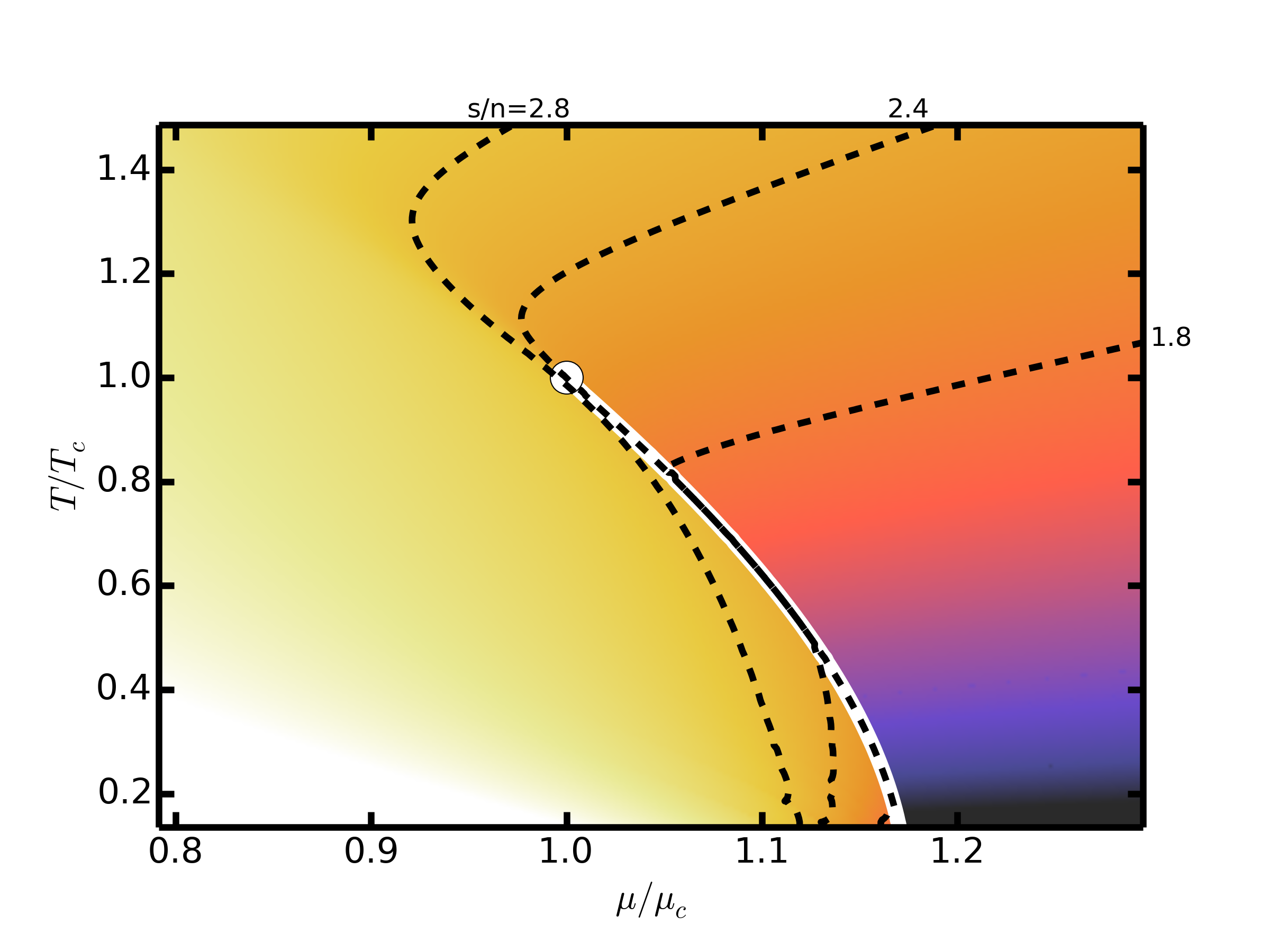}
   \caption{Isentropic curves $s/n$ = const in the scaled $T-\mu$-plane. The first-order phase boundary, depicted by 
            the white curve, terminates in the CEP (white bullet). The dashed white curve is an estimate of the crossover
            region.
            Below the phase transition curve the isentropes are ordered from $s/n = 2.8$ to $s/n = 1.8$ (left to right).
            }
   \label{fig_isentropes}
\end{figure}

In general, local stability requires $s_1 \equiv s(T_c^-) < s_2 \equiv s(T_c^+)$ and $n_1 \equiv n(T_c^-) < n_2\equiv n(T_c^+)$ 
at a first-order
phase transition. However, whether $s_1/n_1 < s_2/n_2$ holds (as for the liquid-gas type transition \cite{Steinheimer:2013xxa}) 
depends on the details of the 
equation of state, \ie the relation of degrees of freedom and latent heat in schematic models. As a 
consequence, the isentropic curves on which \mbox{$s/n$ = const} may bend upward ($s_1/n_1 > s_2/n_2$) or downward 
($s_1/n_1 < s_2/n_2$) on the phase border curve $T_c(\mu)$, supposed $\partial T_c(\mu) / \partial \mu < 0$. 
(For a discussion of isentropes within models superimposing a singular CEP potential and a smooth background in
the spirit of condensed matter approaches, \cf \cite{Bluhm:2006av}). In other words, the phase coexistence curve 
$p_c(T)$ has a positive (negative) slope for the liquid-gas (hadron-quark) type transition, as evidenced by the 
Clapeyron equation $dp_c/dT = (s_1/n_1 - s_2/n_2)/(1/n_1 - 1/n_2)$.
Due to the poor description of baryonic degrees of freedom in the $\LSM$, one must not trust the 
behavior at small temperatures.\footnote{Also other approaches, such as in \cite{Schulze:2009dy} for instance, where hot lattice
data are extrapolated to cold quark star matter, are hampered by a too small pressure at chemical potential in the 
cold nuclear matter region.} Taken as such it is a specific model with a CEP and the liquid-gas transition type
(for a more generic discussion of other models, \cf \cite{Hempel:2013tfa}) with the pattern of isentropic curves as 
exhibited in Fig. \ref{fig_isentropes}. There, the isentrope with $s/n=2.8$ runs slightly below the phase transition line
after bypassing the CEP contrary to those with $s/n<2.8$. These join the $T_c(\mu)$ curve, run some section
along it and depart at a lower "exit temperature". Remarkable for the model with the chosen parameters mentioned
above is the low value of the temperature at the exit point. For other parameter fixings these exit temperatures
can be significantly larger; then also the CEP in absolute $T-\mu$ units changes (see 
\cite{Schaefer:2008hk,Wunderlich:2015rwa} for systematics).

The thermal photon yield is obtained by integrating the photon emission rates along these trajectories from 
certain "initial" points (\cf\cite{Bratkovskaya:2004kv,Ivanov:2005yw} for estimates), 
where the system produced in a heavy-ion collision can be considered as thermalized, until freeze-out, supposed an
adiabatic expansion applies at least approximately.
\section{Photon emissivities}
\begin{figure}
   \centering
   \subfigure{\includegraphics[width=0.16\textwidth,clip]{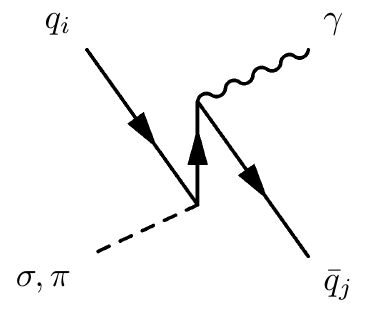}}
   \subfigure{\includegraphics[width=0.16\textwidth,clip]{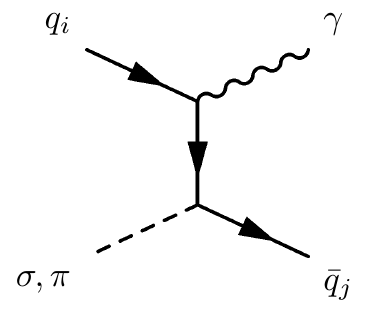}}
   \subfigure{\includegraphics[width=0.16\textwidth,clip]{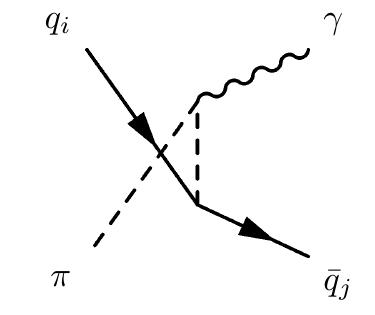}}\\
   \subfigure{\includegraphics[width=0.16\textwidth,clip]{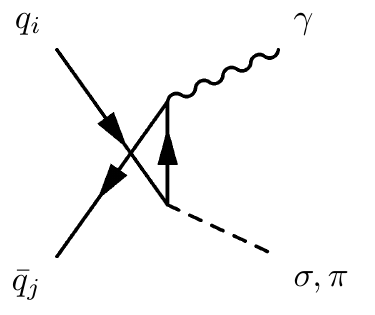}}
   \subfigure{\includegraphics[width=0.16\textwidth,clip]{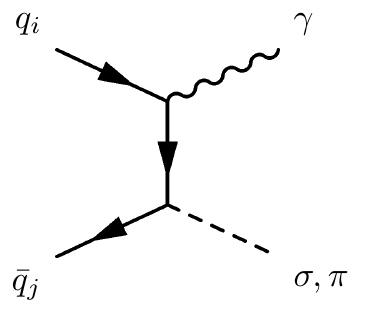}}
   \subfigure{\includegraphics[width=0.16\textwidth,clip]{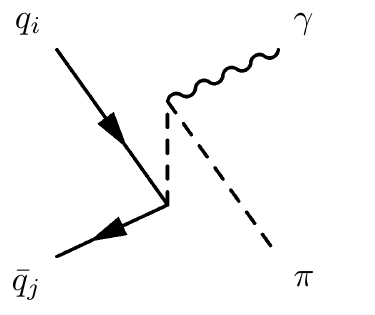}}
   \caption{Lowest-order Feynman diagrams for the processes \eqref{Compton_01}-\eqref{Annihil_01}. 
            The top row shows the diagrams for the Compton processes for quarks and the bottom row
            those for annihilations. If the meson is a $\sigma$ only the left two columns contribute, 
            because the $\sigma$ does not couple to the photon, but (some of) the pions do.}
   \label{fig_Feynman}
\end{figure}

For calculating photon emission rates, the $\LSM$ needs to be equipped with an electromagnetic sector. This is done
by adding a kinetic term 
\begin{equation}
   \Lag_{k\gamma} = \frac14 F^{\mu\nu}F_{\mu\nu}  
\end{equation}
and by coupling the photon field with field strength
tensor $F^{\mu\nu}$ and four-potential $A^\mu$ minimally to quarks
and mesons by replacing $\partial_\mu \rightarrow \partial_\mu + ieQA_\mu$ as done, \eg, in \cite{Fukushima:2010fe} 
for the NJL model.
Considering the fluctuations of the fields $q, \sigma,\bm \pi$ as quanta with effective masses $m^*_{q,\sigma,\pi}$ and
corresponding
dispersion relations $E_i^2 = {m^*_i}^2 + p^2$ the following photon-producing binary reactions in lowest order
\begin{align}
   q_i + \sigma,\pi      &\rightarrow q_j + \gamma        & \text{(Compton scatterings off quarks)},\label{Compton_01}\\
   \bar q_i + \sigma,\pi &\rightarrow \bar q_j + \gamma   & \text{(Compton scat. off antiquarks)},\label{ACompton_01}\\
   q_i + \bar q_j        &\rightarrow \sigma,\pi + \gamma & \text{(annihilations)}\label{Annihil_01}
\end{align}
are accessible within a kinetic theory approach yielding the emission rates
\begin{equation}
 \begin{aligned}
   \omega\frac{d^7N_{12\rightarrow3\gamma}}{dx^4dk^3} 
                               =&\,C\hspace{-0.5mm}\int \frac{d^3 p_1}{2p_1^0}\frac{d^3 p_2}{2p_2^0}\frac{d^3 p_3}{2p_3^0}\delta(p_1+p_2-p_3-k)\\
                                &\times|\mathcal{M}_{12\rightarrow 3\gamma}|^2f_1(p_1)f_2(p_2)\big(1\pm f_3(p_3)\big).
 \end{aligned}\label{rate_01}
\end{equation}
Here $f_i(p_i) = n_{F,B}(p_i^0)$ are the Fermi or Bose distribution functions of the involved species, respectively, 
$\mathcal{M}$ is the corresponding matrix element and $C= (2(2\pi)^8)^{-1}$ is a normalization factor.
In MFA the identification of the effective masses is
\begin{eqnarray}
   m^*_q      &=& g v\\
   {m^*_\sigma}^2 &=& \frac{\partial^2 \Omega_{qq}\big|_{m_q = gv}}{\partial v^2}+\lambda(3v^2 - \zeta),\\
   {m^*_\pi}^2    &=& 2g^2\frac{\partial \Omega_{qq}}{\partial m_q^2}\Bigg|_{m_q = gv} + \lambda(v^2-\zeta),\\
   0          &=& \frac{\partial\Omega_{qq}\big|_{m_q = gv}}   {\partial v} + \lambda(v^3-v\zeta)-H,
\end{eqnarray}
with $v$ being the thermal expectation value of the $\sigma$ field, 
while the linearized fluctuation approach suggests to use
\begin{eqnarray}
   m^*_q      &=& g \left\langle \sqrt{\sigma^2+\vpi^2} \right\rangle,\\
   {m^*_\sigma}^2 &=& \left\langle\frac{\partial^2 \Omega_{qq}}{\partial \Delta^2}\right\rangle
                  + \lambda(3v^2 + 3\langle\Delta^2\rangle + \langle \vpi^2\rangle - \zeta),\\
   {m^*_\pi}^2    &=& \left\langle\frac{\partial^2 \Omega_{qq}}{\partial \pi_a^2}\right\rangle
                  +\lambda(v^2+\langle\Delta^2\rangle+ \frac53\langle\vpi^2\rangle - \zeta),\\
   0          &=& \left\langle\frac{\partial \Omega_{qq}}{\partial \Delta}\right\rangle + \lambda v(v^2+3\langle \Delta^2\rangle + \langle \vpi^2\rangle -\zeta) - H,
\end{eqnarray}
where the variances of the meson fields have to be determined self consistently via 
$\langle \Delta^2\rangle,\langle \pi_a^2 \rangle = 2\partial \Omega_{\sigma,\pi}/\partial (m_{\sigma,\pi_a}^2)$.
Thus the effective masses correspond to thermally averaged "curvature masses" (\cf \cite{Helmboldt:2014iya}). 
The matrix elements squared are calculated according to the Feynman diagrams depicted in Fig. \ref{fig_Feynman}.
The rates \eqref{rate_01} depend on the photon energy $\omega$ and, via the effective masses, 
(see Fig. 1 in \cite{Wunderlich:2014cia}) on temperature and chemical potential 
together with the explicit $T$ and $\mu$ dependence of the $f_i$. 
Given the variations of $m_{q,\sigma,\pi}^*$ over the phase diagram, the photon emission rates \eqref{rate_01} 
map out the phase 
structure \cite{Wunderlich:2014cia,Wunderlich:2015rwa}.
\begin{figure}
   \centering
   \includegraphics[trim=12mm 10mm 10mm 30mm,width=0.48\textwidth,clip]{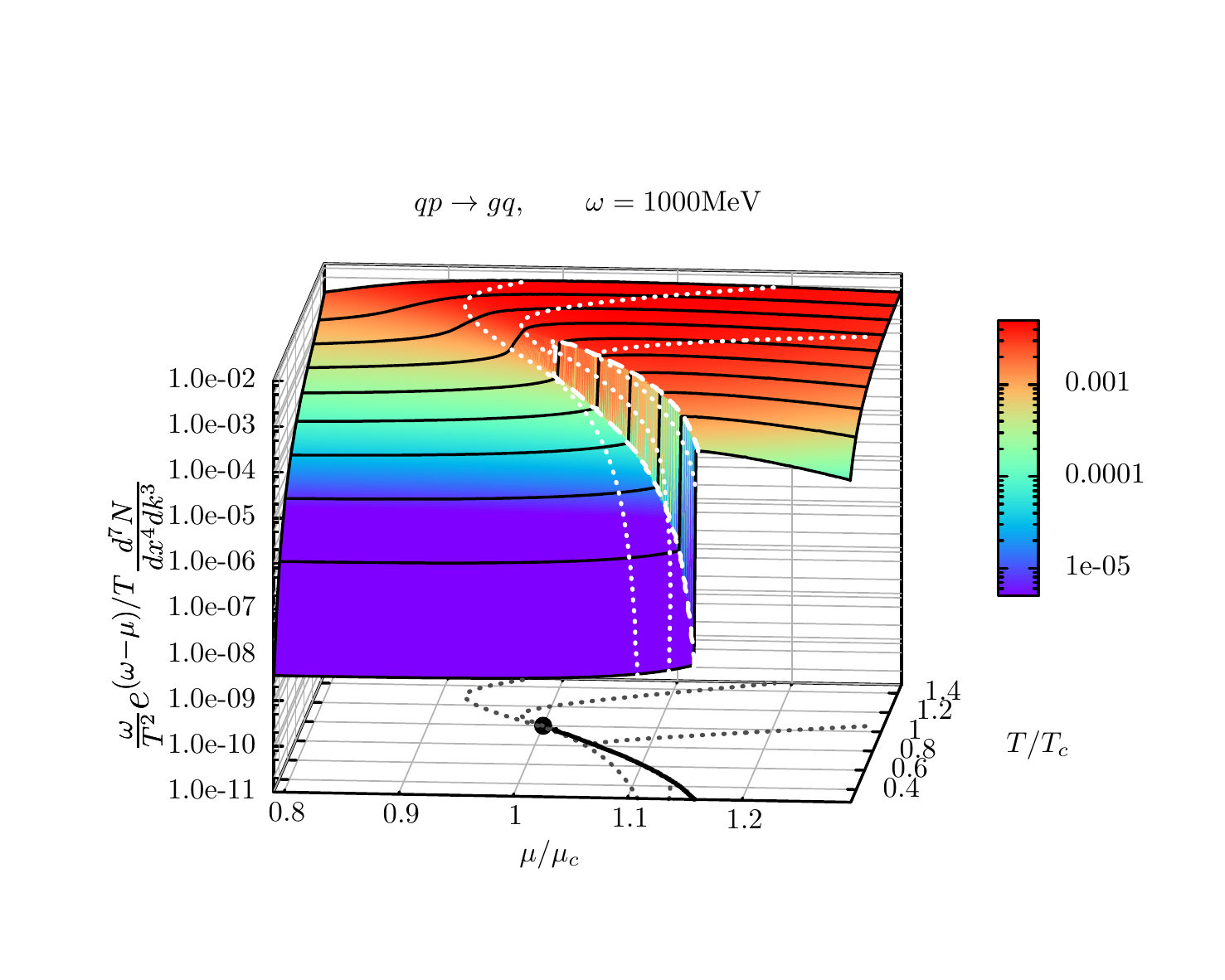}
   \caption{Scaled photon emission rate for $\omega=1\GeV$ in the vicinity of the critical point for the Compton process 
            \eqref{Compton_01} involving $\pi$ mesons. 
            The first-order phase transition curve $T_c(\mu)$ and the CEP are depicted in the bottom plane. Black curves at
            the emission rate surface depict the rate at constant temperature. The dashed white curve surrounds 
            the region where the rate jumps as a consequence of the discontinuity of the masses across the phase
            border line.
            The dotted curves are the rates along the $s/n = 2.8, 2.4, 1.8$ (left to right) 
            isentropes also shown at the base of the plot.}
   \label{fig_Compton_pi}
\end{figure}
\begin{figure}
   \centering
   \includegraphics[trim=5mm 10mm 0mm 30mm,width=0.48\textwidth,clip]{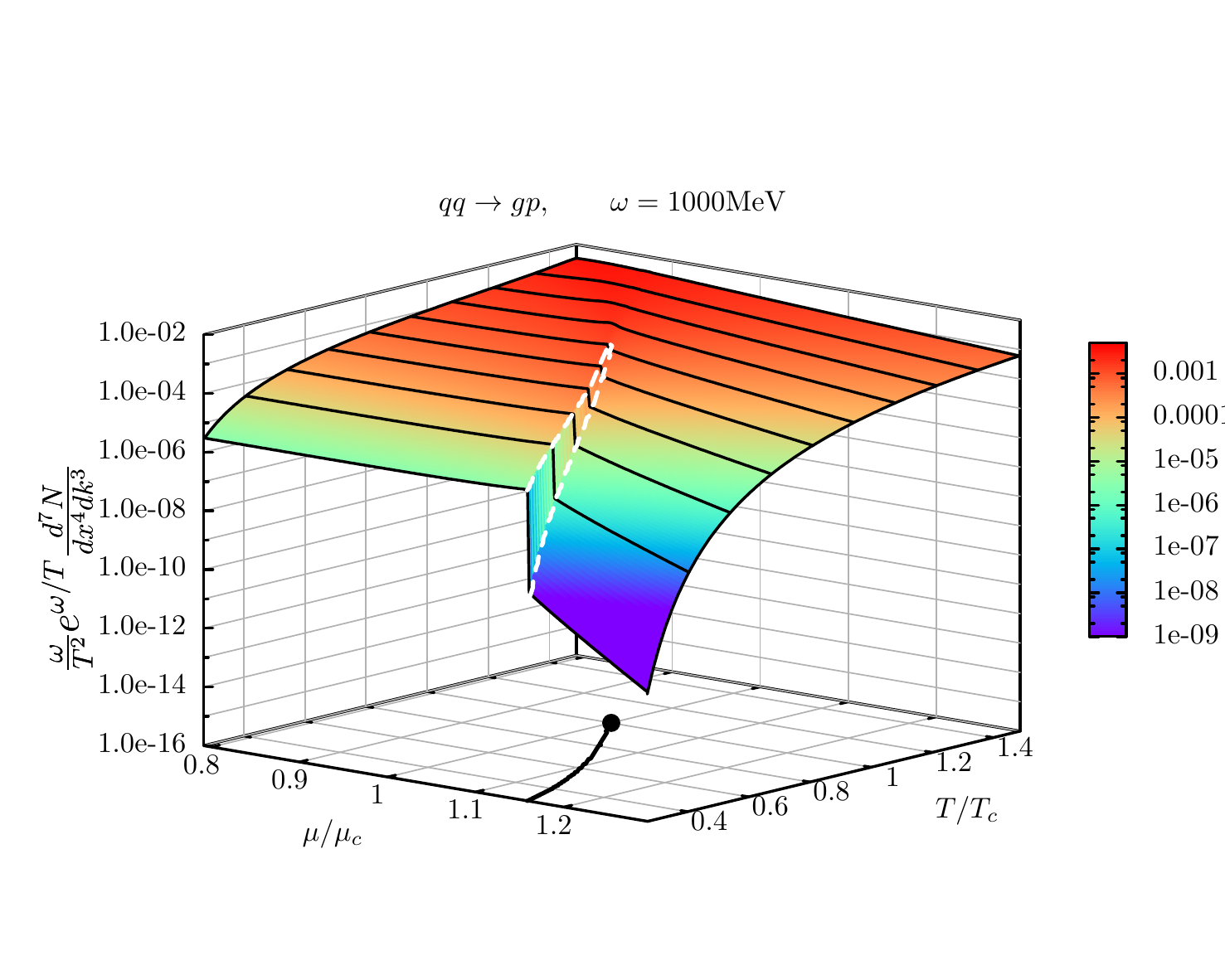}
   \caption{As in Fig. \ref{fig_Compton_pi} but for the annihilation process \eqref{Annihil_01} involving
            $\pi$ mesons. The lines correspond to those of Fig. \ref{fig_Compton_pi}; note that the scaling factor
            differs from the one in Fig. \ref{fig_Compton_pi}.}
   \label{fig_Annihil_pi}
\end{figure}
\begin{figure}
   \centering
   \includegraphics[trim=12mm 10mm 10mm 30mm,width=0.48\textwidth,clip]{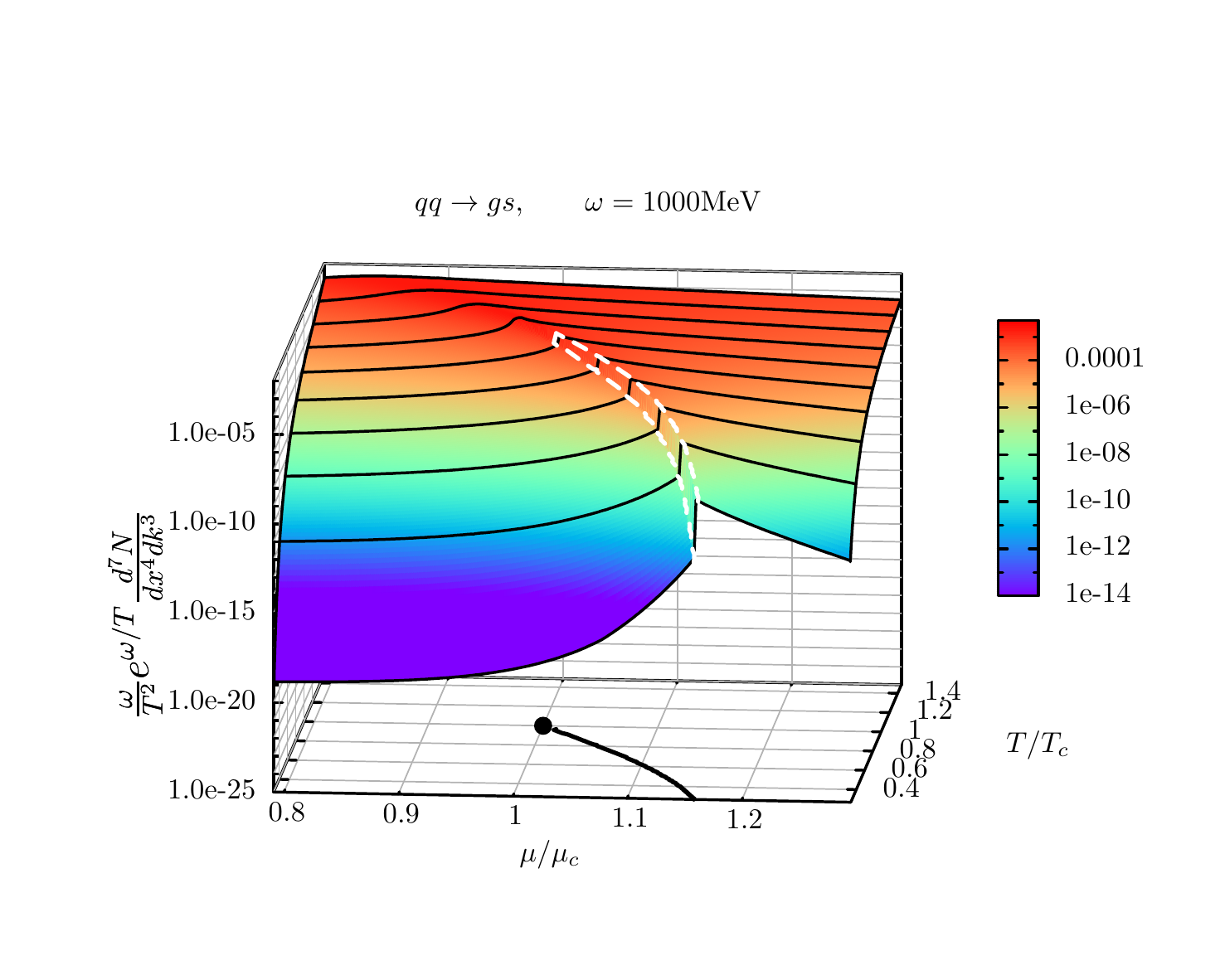}
   \caption{As in Fig. \ref{fig_Compton_pi} but for the annihilation process \eqref{Annihil_01} involving
            $\sigma$ mesons. The lines correspond to those of Fig. \ref{fig_Compton_pi}; scaling factor
            as in Fig. \ref{fig_Annihil_pi}.}
   \label{fig_Annihil_sigma}
\end{figure}
\begin{figure}
   \centering
   \includegraphics[trim=0mm 0mm 20mm 14mm,width=0.48\textwidth,clip]{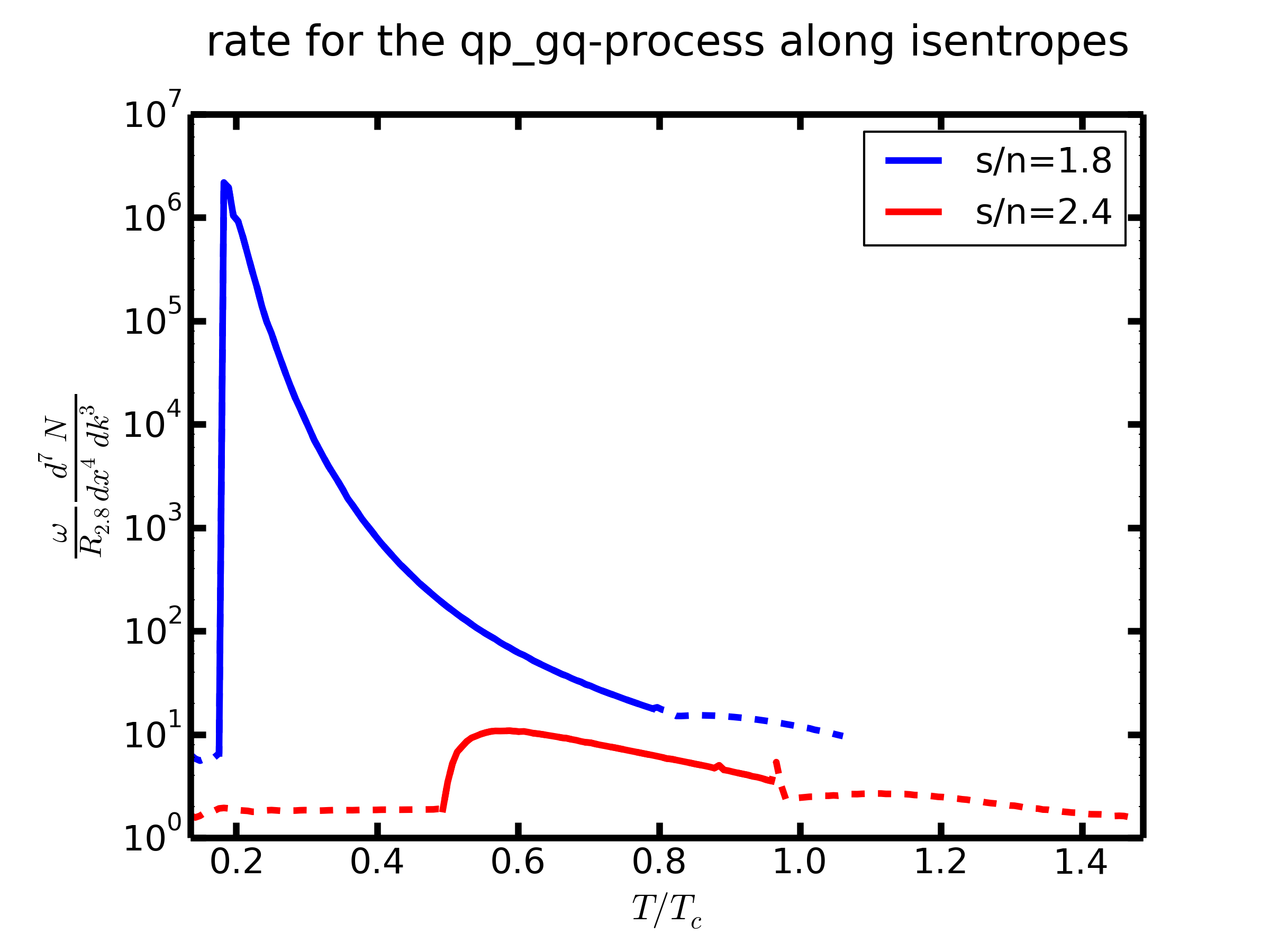}
   \caption{Photon emission rates for the Compton process \eqref{Compton_01} with pions 
            along the isentropes shown in Fig. \ref{fig_isentropes}.
            The rates are scaled by the rate $R_{2.8}$ along the $s/n=2.8$ isentrope for making the 
            influence of the first-order phase transition apparent.
            Solid curves depict the scaled rates in the coexistence region.
            }
   \label{fig_rate_trajectory}
\end{figure}

Selecting $\omega = 1\GeV$ we exhibit in Figs.~\ref{fig_Compton_pi}-\ref{fig_Annihil_sigma} the emission rates for 
the reactions \eqref{Compton_01} (with pions) and \eqref{Annihil_01} (with pions and sigmas). To highlight the 
variations of the emissivities, we display the scaled rates 
$T^{-2} \exp\{(\omega - \mu)/T\} \omega d^7 N_\gamma/(dk^3 dx^4)$, where the exponential scales out the asymptotic 
high-energy behavior \cite{Wunderlich:2015rwa}. Due to the discontinuities of the masses across the phase
transition curve, also the rates reflect these peculiarities: Depending on the participating species, the 
respective rates may jump up as in Figs. \ref{fig_Compton_pi} and \ref{fig_Annihil_sigma} or 
drop down as in Fig. \ref{fig_Annihil_pi} at the phase boundary when crossing it at constant temperature
with increasing chemical potential. The size of the jumps as well as their direction depends mostly on the 
mass of the outgoing particle (besides the photon). Within a Boltzmann approximation for the distribution functions
it can be shown that for large $\omega$ the dominant mass dependence is $\propto \exp\{-m_3/T\}$ \cite{Wunderlich:2015rwa}. 
Thus, the decrease of the quark and sigma mass (for constant temperature and increasing chemical potential) leads 
to an increase of the Compton rate (\cf Fig. \ref{fig_Compton_pi}) or the annihilation rate with sigmas 
(\cf Fig. \ref{fig_Annihil_sigma}), while an increase of the pion mass leads to a dropping rate in the 
annihilation case with pions (\cf Fig. \ref{fig_Annihil_pi}). The size of the jumps changes from one order of magnitude 
near the CEP to many orders of magnitude for small $T$ depending also on the process under consideration. 
Beyond the CEP and the first-order phase transition line the rates vary smoothly over the $T-\mu$-plane.

To arrive at some impression of the variation of the rates along an adiabatic expansion path, we exhibit
in Fig.~\ref{fig_rate_trajectory} the rates along isentropic trajectories in the vicinity of the phase border curve.
The solid curves are for the scaled rates, where the isentropic curve runs on the coexistence curve $T_c(\mu)$.
Here, the rates are a superposition in the spirit of a two-phase mixture, \ie 
$dN = dN_+(T^+(\mu)) x + dN_-(T^-(\mu))(1-x)$, where x means the weight of the high-temperature phase.
(Analog constructions of two-phase mixtures apply for density, entropy density and energy density).
The large ratios of the rates depicted in Fig.~\ref{fig_rate_trajectory} for the $s/n=1.8$ curve can be 
understood in terms of phase mixing in the coexistence region. 
At the phase boundary the photon emission originates from both phases. 
For the Compton-process the above mentioned leading mass dependence is $\propto \exp\{-m_q/T\}$. Because the drop of 
the quark mass is of order $200\MeV$ at small temperatures (\cf Fig. 1c in \cite{Wunderlich:2014cia}) the rate in
the high-temperature phase is enhanced w.r.t. the low-temperature phase emission rate 
by a factor of $\exp\{\Delta m_q/T\}$ which can achieve values up to $10^5$ close to the exit temperature of
these isentropes, even for small phase weight of the high-temperature phase.
For the rates along the isentropes depicted in Figs.~\ref{fig_isentropes} and \ref{fig_Compton_pi}, this means 
that these curves stay close to the
upper rim of the discontinuity even if $T$ is only slightly larger than the above mentioned "exit temperature" 
(\cf Fig.~\ref{fig_Compton_pi}) and
the huge enhancement factor corresponds to a jump of the rates of the same order of magnitude.
Despite these huge differences in the emission rates at temperatures considerably smaller than the CEP temperature,
the total photon yield is less wide spread, because the major part of the thermal photon emission is from the 
hot medium with $T>T_c(\mu_c)$, where the isentropic emission rates in Fig.~\ref{fig_rate_trajectory} are of the 
same order of magnitude.

Clearly, besides the $2\rightarrow2$ processes considered so far, the other channels must be accounted for 
to arrive at firm conclusions.
Nevertheless, these results support the hypotheses that the photon emission may obey distinctive peculiarities
when the matter in the course of a heavy-ion collision undergoes a first-order phase transition, supposed it is 
strong enough.
\section{Summary}
In summary, we have shown that a few selected channels for the photon emissivity directly map out the phase diagram.
Despite of the great progress of ab initio calculations for a variety of quantities characterizing
the properties of strongly interacting matter,
the region of non-zero baryon density is less reliable accessible in a wide range.
Therefore, we resort here to a specific model which displays a critical end point (CEP), where a curve of first-order
phase transitions, which are of the liquid-gas type, sets in and continues to larger baryon densities.\footnote{Additionally to the above disclaimers, we
stress that we do not address the CEP itself, but focus on the emerging first-order transition effects. 
A crucial issue is of course the actual location of the CEP and whether in heavy-ion collisions
values of $T$ and $\mu$ are accessible, which are beyond the phase transition curve.}
The considered photon generating processes refer to lowest-order reactions
of the involved effective degrees of freedom.
For the corresponding rates a crucial ingredience is represented by the so-called derivative masses which obey significant
variations, or even discontinuities, in the vicinity of the critical endpoint, causing a strong imprint on the rates.
Clearly, the consideration of many more channels along with improved inclusion of QCD like degrees of freedom is necessary to arrive at
realistic estimates of the total emission rate. Equally challenging is also the account of the dynamics
(\cf \cite{Wunderlich:2012db} for a method to follow the smoothed longitudinal evolution of matter or 
\cite{Steinheimer:2013xxa} for the treatment of spinodal clumping in the mixed phase).
Nevertheless, we hope that our explorative investigations have digged out pertinent features of modifications of
penetrating probes monitoring the evolution of matter in the course of heavy-ion collisions in the energy range of NICA.
Guided by the experience from macroscopic examples of critical point phenomena, \eg critical opalescence emerging from a 
diverging correlation length in approaching the CEP, one would expect more drastic signatures, \eg related to a strong 
spike in the specific heat or a stall/push of the expansion dynamics, which in turn would shape the space-time integrated
photon spectra. In our approach, however, the CEP imprints on the net-photon emission rate is rather modest due to the 
solely coupling to the derivative masses of effective degrees of freedom.
\section{Acknowledgments}
One of the authors (BK) gratefully acknowledges useful conversations with J. Randrup on the relation of liquid-gas
and hadron-quark transitions. The work is supported by BMBF grant 05P12CRGH.
\bibliographystyle{aip}
\bibliography{nica}
\end{document}